\documentclass[aps,prb,reprint,superscriptaddress]{revtex4-2}
\usepackage[T1]{fontenc}
\usepackage{lmodern}
\usepackage{graphicx}
\usepackage{amsmath}
\usepackage{amssymb}
\usepackage{color}
\usepackage{hyperref}

\begin{document}

\title{Stacking-ratio-tuned interlayer coupling in $\text{EuCu}_2\text{ZnAs}_2$ and $\text{Eu}_3\text{Cu}_4\text{Zn}_4\text{As}_6$ intergrowths}

\author{Zhiyu Zhou}
\affiliation{Key Laboratory of Quantum Materials and Devices of Ministry of Education, School of Physics, Southeast University, Nanjing 211189, China}
\author{Xiyu Chen}
\affiliation{Key Laboratory of Quantum Materials and Devices of Ministry of Education, School of Physics, Southeast University, Nanjing 211189, China}
\author{Xin-Yu Zhao}
\affiliation{School of Physics, Interdisciplinary Center for Quantum Information and State Key Laboratory of Silicon and Advanced Semiconductor Materials, Zhejiang University, Hangzhou 310058, China}
\author{Yaran Duan}
\affiliation{Key Laboratory of Quantum Materials and Devices of Ministry of Education, School of Physics, Southeast University, Nanjing 211189, China}
\author{Shiran Tao}
\affiliation{Key Laboratory of Quantum Materials and Devices of Ministry of Education, School of Physics, Southeast University, Nanjing 211189, China}
\author{Guang-Han Cao}
\affiliation{School of Physics, Interdisciplinary Center for Quantum Information, and State Key Laboratory of Silicon and Advanced Semiconductor Materials, Zhejiang University, Hangzhou 310058, China}
\author{Shuai Dong}
\affiliation{Key Laboratory of Quantum Materials and Devices of Ministry of Education, School of Physics, Southeast University, Nanjing 211189, China}
\author{Zhi-Cheng Wang}
\email{wzc@seu.edu.cn}
\affiliation{Key Laboratory of Quantum Materials and Devices of Ministry of Education, School of Physics, Southeast University, Nanjing 211189, China}

\date{\today}

\begin{abstract}

Function-oriented materials design increasingly relies on modular construction. Using a bottom-up block-layer (BL) strategy, two layered intergrowth compounds, $\text{EuCu}_2\text{ZnAs}_2$ (1212 phase) and $\text{Eu}_3\text{Cu}_4\text{Zn}_4\text{As}_6$ (3446 phase), are designed and grown from Pb flux. Both structures stack alternating Zintl-type $[\text{EuZn}_2\text{As}_2]$ and metallic $[\text{EuCu}_4\text{As}_2]$ blocks along the $c$-axis and retain the rhombohedral symmetry ($R\bar{3}m$) of the Cu-rich parent. Varying the BL stacking ratio from $1:1$ to $1:2$ tunes interlayer magnetic coupling and carrier density. $\text{EuCu}_2\text{ZnAs}_2$ orders antiferromagnetically at $T_{\text{N}} = 22.5\text{ K}$ and, on further cooling, undergoes an AFM-to-AFM spin reorientation, whereas $\text{Eu}_3\text{Cu}_4\text{Zn}_4\text{As}_6$ enters a canted AFM state at $T_{\text{N}} = 20.5\text{ K}$ that carries a weak ferromagnetic moment arising from inequivalent Eu sublattices. Both compounds show metallic transport dominated by the $[\text{EuCu}_4\text{As}_2]$ sublattices, with negative magnetoresistance above $T_{\text{N}}$ that reverses to a large positive MR in the ordered state; the extra Zintl spacer in the 3446 phase dilutes the carrier concentration and lowers the conductivity. These results establish the block-layer stacking ratio as a direct, disorder-free control parameter for tuning magnetic order and carrier density, and point to a general route to functional magnetic intergrowths.

\end{abstract}
\maketitle

\section{Introduction}

Modern condensed-matter physics and materials chemistry increasingly rely on function-oriented design to realize novel quantum states and targeted macroscopic properties \cite{dagotto2005,keimer2017}. Instead of chemical substitution within a single parent lattice, a bottom-up approach assembles pre-functionalized structural modules, such as two-dimensional layers or block layers (BLs), into heterostructured intergrowth architectures along a specific crystallographic axis \cite{tokura1990,jansen2002a,geim2013}. In these ``Lego-like'' superlattices the building blocks largely keep their intrinsic physical character, while interfacial coupling, strain, and charge transfer between dissimilar sublattices produce emergent phenomena absent in the parent phases \cite{novoselov2016,ohtomo2004,hwang2012,rao1985}. Layered superconductors illustrate the idea: combining distinct block layers generates broad families of superconducting intergrowths without external chemical doping \cite{Sun2012Ba2Ti2Fe2As4OANewSuperconductorContainingFe2As2LayersandTi2OSheets,iyo2016,Liu2016SuperconductivityandFerromagnetisminHoleDopedRbEuFe4As4,Wang2016SuperconductivityinKCa2Fe4As4F2withSeparateDoubleFe2As2Layers,li2025,li2025a}.

The physical properties of intergrowth materials are highly sensitive to stacking geometry, including spatial sequence, interlayer distance, and unit-cell stacking ratios \cite{geim2013}. In two-dimensional van der Waals heterostructures and their bulk intergrowth analogues, small changes in stacking symmetry or layer thickness reshape magnetic order, electronic topology, and transport \cite{burch2018}. The magnetic topological insulator series $\mathrm{MnBi_2Te_4}\cdot n(\mathrm{Bi_2Te_3})$ ($n = 0, 1, 2, \dots$) shows this clearly: inserting non-magnetic $\mathrm{Bi_2Te_3}$ quintuple-layer spacers tunes the interlayer exchange coupling from three-dimensional antiferromagnetism to decoupled two-dimensional ferromagnetism, driving transitions among axion-insulator, Chern-insulator, and Weyl-semimetal states \cite{otrokovPredictionObservationAntiferromagnetic2019,dengQuantumAnomalousHall2020,huRecentProgressMnBi22024,liProgressAntiferromagneticTopological2024}. Precise control of the BL stacking ratio is therefore a direct route to tuning macroscopic physical properties.

Layered pnictides are well suited to this modular construction. Zintl-type 122 phases such as $\mathrm{EuZn_2As_2}$ ($\mathrm{CaAl_2Si_2}$-type, $P\bar{3}m1$) show $A$-type antiferromagnetic (AFM) order ($T_{\text{N}} \approx 19\text{ K}$) together with magnetic topological configurations \cite{klufersAB2X2VerbindungenImCaAl2Si2Typ1980,Wang2022Anisotropyofthemagneticandtransportproperties}. Copper-rich 142 phases such as $\mathrm{EuCu_4As_2}$ ($\mathrm{CaCu_4P_2}$-type, $R\bar{3}m$) are metallic, with ferromagnetic (FM) correlations ($T_{\text{C}} \approx 34\text{ K}$) and large magnetoresistance (MR) \cite{Dunner1999SynthesisandcrystalstructureofACu4As2A,Li2022Largepositiveandnegativemagnetoresistanceinthemagnetic,Nie2023Subtlelatticedistortiondrivenphasetransitionsinlayered}. Structural relatives across the 142 family, including $\mathrm{EuAg_4Sb_2}$, $\mathrm{SrCu_4As_2}$, and $\mathrm{CaCu_4As_2}$, show a range of states such as spin-Moir\'e textures from noncollinear magnetic order, linear MR, and multilayer quantum Hall physics \cite{greenRobustSpinMoireSuperlatticeDrivenGap2025,greenMappingThreedimensionalFermiology2025a,shenStructuralDistortionIncommensurate2020,neves2026,neves2026a,malickQuantumOscillationsTransport2025,sasmal2022,malickLargeMagnetoresistanceFirstorder2024}. Because $\mathrm{EuZn_2As_2}$ ($a = 4.211\text{ \AA}$) and $\mathrm{EuCu_4As_2}$ ($a = 4.183\text{ \AA}$) differ by less than 1\% in-plane and share matching triangular $\mathrm{Eu^{2+}}$ sublattices, the two families can be interlinked directly to engineer intergrowths.

This work applies the strategy to two representative intergrowths, $\mathrm{EuCu_2ZnAs_2}$ (1212 phase) and $\mathrm{Eu_3Cu_4Zn_4As_6}$ (3446 phase). $\mathrm{EuCu_2ZnAs_2}$ is a strict $1:1$ alternating sequence of metallic $[\mathrm{EuCu_4As_2}]$ and Zintl $[\mathrm{EuZn_2As_2}]$ BLs, whereas $\mathrm{Eu_3Cu_4Zn_4As_6}$ adopts a $1:2$ sequence with an extra Zintl spacer per repeat unit. Varying the BL stacking ratio controls interfacial charge transfer, carrier density, and interlayer exchange coupling. Inserting the extra Zintl layer in the 3446 phase dilutes the carrier concentration and breaks local inversion symmetry across inequivalent Eu sites, producing a canted weak FM moment that the 1212 phase lacks. Because $\mathrm{CaAl_2Si_2}$- and $\mathrm{CaCu_4P_2}$-type structures are widespread, the block-layer strategy should extend to large families of 1212- and 3446-type intergrowths through chemical substitution.


\section{Experimental Section}
\subsection{Crystal growth}

Single crystals of $\text{EuCu}_2\text{ZnAs}_2$ and $\text{Eu}_3\text{Cu}_4\text{Zn}_4\text{As}_6$ were grown by a high-temperature Pb-flux method from high-purity Eu pieces (99.9\%), Zn powder (99.999\%), Cu powder (99.9\%), As lumps (99.99\%), and Pb shots (99.999\%). Before reaction, the oxide layer on the Eu pieces was scraped off with wire cutters. All reagents and precursors were handled in an Ar-filled glovebox with $\text{H}_2\text{O}$ and $\text{O}_2$ levels below 0.01 ppm to prevent oxidation or hydrolysis.

The synthesis was carried out in two steps. First, binary precursors $\text{EuAs}$ and $\text{Cu}_2\text{As}$ were prepared separately. Stoichiometric mixtures of $\text{Eu}:\text{As} = 1:1$ and $\text{Cu}:\text{As} = 2:1$ were loaded into alumina crucibles and sealed in evacuated silica tubes under high vacuum. The tubes were heated to 400 $^\circ\text{C}$ over 400 min, held at 400 $^\circ\text{C}$ for 1 day, heated to 900 $^\circ\text{C}$ over 400 min, held for 3 days, and then cooled naturally to room temperature. The polycrystalline products were ground into fine powders.

In the second step, the $\text{EuAs}$ and $\text{Cu}_2\text{As}$ powders were mixed with Zn powder in a molar ratio $\text{EuAs}:\text{Cu}_2\text{As}:\text{Zn} = 1:1:1$. About 0.5 g of this mixture, with 2 g of Pb shots, was loaded into an alumina crucible and sealed in an evacuated silica tube. Single crystals of the 1212 phase ($\text{EuCu}_2\text{ZnAs}_2$) were synthesized by heating the sealed tube to 500 $^\circ\text{C}$ over 500 min, holding for 1 day, ramping to 1100 $^\circ\text{C}$ over 500 min, and holding for 10 h. The tube was then cooled to 800 $^\circ\text{C}$ at 3 $^\circ\text{C/h}$, followed by high-temperature centrifugation to separate the excess molten Pb flux.

Single crystals of the 3446 phase ($\text{Eu}_3\text{Cu}_4\text{Zn}_4\text{As}_6$) were grown with identical precursor stoichiometry and loading, with phase selection set by adjustments to the growth window and thermal history. The mixture was heated to 1050 $^\circ\text{C}$ (instead of 1100 $^\circ\text{C}$) and slowly cooled to 850 $^\circ\text{C}$ at 1 $^\circ\text{C/h}$ before centrifugation. The slower cooling stabilizes the $[\text{EuCu}_4\text{As}_2]:[\text{EuZn}_2\text{As}_2] = 1:2$ stacking sequence, presumably by allowing the repetitive assembly of $[\text{EuZn}_2\text{As}_2]$ BLs. Despite the near-identical conditions, each optimized batch yielded essentially single-phase crystals, with no discernible intermixing between the 1212 and 3446 phases. The crystals are air-stable under ambient conditions.

\subsection{Structure and chemical characterization}

Suitable single crystals of $\text{EuCu}_2\text{ZnAs}_2$ and $\text{Eu}_3\text{Cu}_4\text{Zn}_4\text{As}_6$ were selected for single-crystal X-ray diffraction (SCXRD). Intensity data were collected at 300(2)~K on an XtaLAB Synergy R four-circle diffractometer with a HyPix detector and a Mova (Mo) micro-focus sealed X-ray tube with a mirror monochromator (Mo K$\alpha$ radiation, $\lambda$ = 0.71073 \AA). Data reduction and empirical absorption corrections were performed with the CrysAlisPro software suite. The structures were solved by intrinsic phasing with SHELXT and refined by full-matrix least-squares on $F^2$ with SHELXL in the Olex2 graphical interface \cite{Olex2, ShelXle}. Anisotropic atomic displacement parameters were refined for all atoms. Data collection parameters, index limits ($h,k,l$), $2\theta$ ranges, and full refinement statistics are given in Table S1 of the Supplemental Material (SM) \cite{SM}. Structural visualizations were generated with the VESTA software package \cite{Vesta}.

Energy-dispersive X-ray spectroscopy (EDS) was performed on an FEI Inspect F50 scanning electron microscope (SEM) equipped with an EDAX detector to determine chemical compositions and surface microstructures.

\subsection{Physical property measurements}
Direct-current (DC) magnetization, heat capacity, and electrical transport for both compounds were measured on a Quantum Design Physical Property Measurement System (PPMS DynaCool). Additional magnetic measurements were acquired on a Quantum Design Magnetic Property Measurement System (MPMS3). Electrical resistivity was measured with a standard four-probe configuration.

\section{Results and Discussion}
\subsection{Crystal structure design, synthesis, and intergrowth phase stability}

\begin{figure*}
	\centering
	\includegraphics[width=0.95\textwidth]{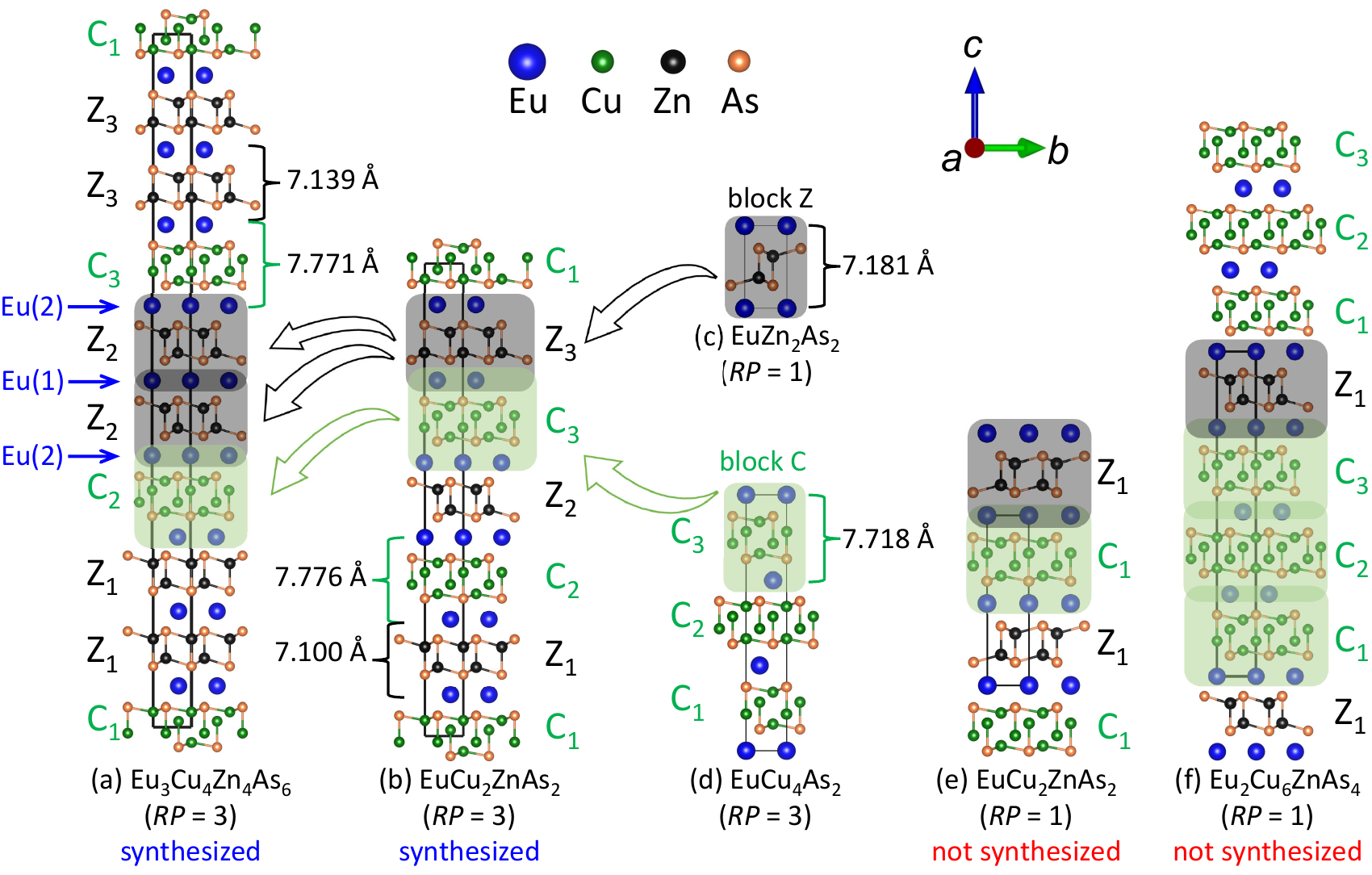}
	\caption{Crystal structures of the two intergrowth compounds (a) $\text{Eu}_3\text{Cu}_4\text{Zn}_4\text{As}_6$ and (b) $\text{EuCu}_2\text{ZnAs}_2$, their parent phases (c) $\text{EuZn}_2\text{As}_2$ and (d) $\text{EuCu}_4\text{As}_2$, and two designed but unrealized structures (e, f). The [$\text{EuZn}_2\text{As}_2$] and [$\text{EuCu}_4\text{As}_2$] BLs are highlighted by semi-transparent gray and green rectangles, respectively. Eu, Cu, Zn, and As atoms are shown as blue, black, green, and yellow spheres, respectively. All structures are viewed along the crystallographic $a$ axis.}
	\label{F1}
\end{figure*}

Using a bottom-up structural engineering strategy, we designed and synthesized two heterostructured intergrowth compounds, $\text{EuCu}_2\text{ZnAs}_2$ (Fig.~\ref{F1}(b)) and $\text{Eu}_3\text{Cu}_4\text{Zn}_4\text{As}_6$ (Fig.~\ref{F1}(a)), by alternately stacking the building blocks of the two parent phases, $\text{EuZn}_2\text{As}_2$ and $\text{EuCu}_4\text{As}_2$, along the crystallographic $c$-axis. Both parents consist of triangular $\text{Eu}^{2+}$ sheets interleaved with distinct polyhedral sublattices. $\text{EuZn}_2\text{As}_2$ crystallizes in the $\text{CaAl}_2\text{Si}_2$-type structure ($P\bar{3}m1$, No.~164) with a single $[\text{EuZn}_2\text{As}_2]$ BL per unit cell (repeat period $RP = 1$, light gray in Fig.~\ref{F1}(c)). In contrast, $\text{EuCu}_4\text{As}_2$ adopts the $\text{CaCu}_4\text{P}_2$-type structure ($R\bar{3}m$, No.~166) with three identical $[\text{EuCu}_4\text{As}_2]$ BLs ($RP = 3$, light green in Fig.~\ref{F1}(d)), related by the rhombohedral centering vector $(2/3, 1/3, 1/3)$ and designated $C_1$, $C_2$, and $C_3$.

Given the structural compatibility and the sub-1\% in-plane lattice mismatch between $\text{EuZn}_2\text{As}_2$ ($a = 4.211$~\AA) and $\text{EuCu}_4\text{As}_2$ ($a = 4.183$~\AA) at ambient conditions \cite{Dunner1999SynthesisandcrystalstructureofACu4As2A,Wang2022Anisotropyofthemagneticandtransportproperties}, coherent intergrowth along shared $\text{Eu}^{2+}$ planes is favorable. We considered four candidate stacking configurations:
(i) $1 \times [\text{EuCu}_4\text{As}_2] + 2 \times [\text{EuZn}_2\text{As}_2]$ BLs per repeat unit with $RP = 3$ ($-C_1Z_1Z_1-C_2Z_2Z_2-C_3Z_3Z_3-$ stacking), giving $\text{Eu}_3\text{Cu}_4\text{Zn}_4\text{As}_6$ (Fig.~\ref{F1}(a));
(ii) $1 \times [\text{EuCu}_4\text{As}_2] + 1 \times [\text{EuZn}_2\text{As}_2]$ BLs with $RP = 3$ ($-C_1Z_1-C_2Z_2-C_3Z_3-$ stacking), giving $\text{EuCu}_2\text{ZnAs}_2$ (Fig.~\ref{F1}(b));
(iii) an alternative $1:1$ stacking variant with $RP = 1$ ($-C_1Z_1-$ sequence, Fig.~\ref{F1}(e)); and
(iv) a $3:1$ configuration with $RP = 1$ ($-C_1C_2C_3Z_1-$ sequence, nominal formula $\text{Eu}_2\text{Cu}_6\text{ZnAs}_4$, Fig.~\ref{F1}(f)).

\begin{table*}
	\caption{Crystallographic data and structure refinement parameters for $\text{EuCu}_2\text{ZnAs}_2$ and $\text{Eu}_3\text{Cu}_4\text{Zn}_4\text{As}_6$ collected at 300~K (Mo K$\alpha$ radiation, $\lambda = 0.71073$~\AA), in comparison with the parent phases \cite{Dunner1999SynthesisandcrystalstructureofACu4As2A,Wang2022Anisotropyofthemagneticandtransportproperties}.}
	\label{T1}
	\centering
	\begin{tabular}{lllll}
		\hline
		Formula & $\text{EuCu}_2\text{ZnAs}_2$ & $\text{Eu}_3\text{Cu}_4\text{Zn}_4\text{As}_6$ & $\text{EuCu}_4\text{As}_2$ & $\text{EuZn}_2\text{As}_2$ \\
		\hline
		Crystal system & Trigonal & Trigonal & Trigonal & Trigonal \\
		Space group & $R\bar{3}m$ (No. 166) & $R\bar{3}m$ (No. 166) & $R\bar{3}m$ (No. 166) & $P\bar{3}m1$ (No. 164) \\
		$a$ (\AA) & 4.2024(4) & 4.2028(1) & 4.183(1) & 4.21118(3) \\
		$c$ (\AA) & 44.626(5) & 66.144(3) & 23.154(2) & 7.18114(6) \\
		$V$ (\AA$^3$) & 682.52(15) & 1011.81(6) & 350.86(5) & 110.2888(24) \\
		$Z$ & 6 & 3 & 3 & 1 \\
		$F_\mathrm{w}$ (g/mol) & 494.10 & 1421.04 & 555.99 & 432.57 \\
		$\rho_\mathrm{calc}$ (g/cm$^3$) & 7.215 & 6.996 & 7.894 & 6.510 \\
		$R_1$ [$F^2 > 2\sigma(F^2)$]\textsuperscript{\emph{a}} & 3.21\% & 4.83\% & -- & -- \\
		$wR_2$ (all data)\textsuperscript{\emph{b}} & 8.58\% & 11.53\% & -- & -- \\
		GOF & 1.240 & 1.097 & -- & -- \\
		\hline
	\end{tabular}

	\begin{tabular}{llllllllll}
		\hline
		Atom & Site & $x$ & $y$ & $z$ & Occ. & $U_\mathrm{eq}$ (\text{\AA}$^2 \times 10^3$)\textsuperscript{\emph{c}} & $U_{11}$\textsuperscript{\emph{d}} & $U_{33}$ & $U_{12}$ \\
		\hline
		\multicolumn{10}{l}{\textit{$\text{EuCu}_2\text{ZnAs}_2$ (1212 phase)}} \\
		Eu & 6c & 0 & 0 & 0.42045(2) & 1.00 & 9.6(3) & 10.3(3) & 8.1(4) & 5.17(15) \\
		Cu(1) & 6c & 0 & 0 & 0.34584(3) & 0.985(7) & 20.4(5) & 24.9(5) & 11.3(7) & 12.5(3) \\
		Cu(2) & 6c & 0 & 0 & 0.03230(3) & 0.983(7) & 17.5(5) & 14.5(5) & 23.6(8) & 7.2(3) \\
		Zn & 6c & 0 & 0 & 0.18741(3) & 0.993(7) & 12.5(4) & 12.9(5) & 11.8(6) & 6.5(3) \\
		As(1) & 6c & 0 & 0 & 0.29085(2) & 1.00 & 9.7(3) & 10.5(3) & 7.9(4) & 5.27(17) \\
		As(2) & 6c & 0 & 0 & 0.12911(2) & 1.00 & 8.4(3) & 8.7(3) & 7.9(5) & 4.37(17) \\
		\hline
		\multicolumn{10}{l}{\textit{$\text{Eu}_3\text{Cu}_4\text{Zn}_4\text{As}_6$ (3446 phase)}} \\
		Eu(1) & 3b & 0 & 0 & 0.50000 & 1.00 & 8.5(4) & 6.7(4) & 12.3(6) & 3.3(2) \\
		Eu(2) & 6c & 0 & 0 & 0.39208(2) & 0.998(3) & 8.9(3) & 7.7(4) & 11.2(5) & 3.84(18) \\
		Cu(1) & 6c & 0 & 0 & 0.02184(3) & 0.982(6) & 16.5(6) & 11.9(7) & 25.7(12) & 5.9(4) \\
		Cu(2) & 6c & 0 & 0 & 0.34188(3) & 1.00 & 20.4(5) & 22.9(7) & 15.5(10) & 11.4(3) \\
		Zn(1) & 6c & 0 & 0 & 0.12662(3) & 1.00 & 11.7(4) & 10.0(6) & 15.3(9) & 5.0(3) \\
		Zn(2) & 6c & 0 & 0 & 0.23493(3) & 0.984(5) & 10.6(5) & 9.4(7) & 12.8(9) & 4.7(3) \\
		As(1) & 6c & 0 & 0 & 0.19559(2) & 1.00 & 7.2(4) & 5.3(5) & 10.9(8) & 2.6(3) \\
		As(2) & 6c & 0 & 0 & 0.08696(2) & 1.00 & 7.6(4) & 5.5(5) & 11.8(8) & 2.7(2) \\
		As(3) & 6c & 0 & 0 & 0.30470(2) & 1.00 & 9.1(4) & 7.3(5) & 12.6(7) & 3.7(3) \\
		\hline
	\end{tabular}

	\textsuperscript{\emph{a}} $R_1 = \Sigma||F_o| - |F_c|| / \Sigma|F_o|$.
	\textsuperscript{\emph{b}} $wR_2 = [\Sigma w(F_o^2 - F_c^2)^2 / \Sigma w(F_o^2)^2]^{1/2}$.
	\textsuperscript{\emph{c}} $U_\mathrm{eq}$ is defined as one-third of the trace of the orthogonalized $U_{ij}$ tensor.
	\textsuperscript{\emph{d}} $U_{22}=U_{11}$, $U_{12}=0.5U_{11}$, and $U_{13}=U_{23}=0$.

\end{table*}

\begin{figure}
	\centering
	\includegraphics[width=\columnwidth]{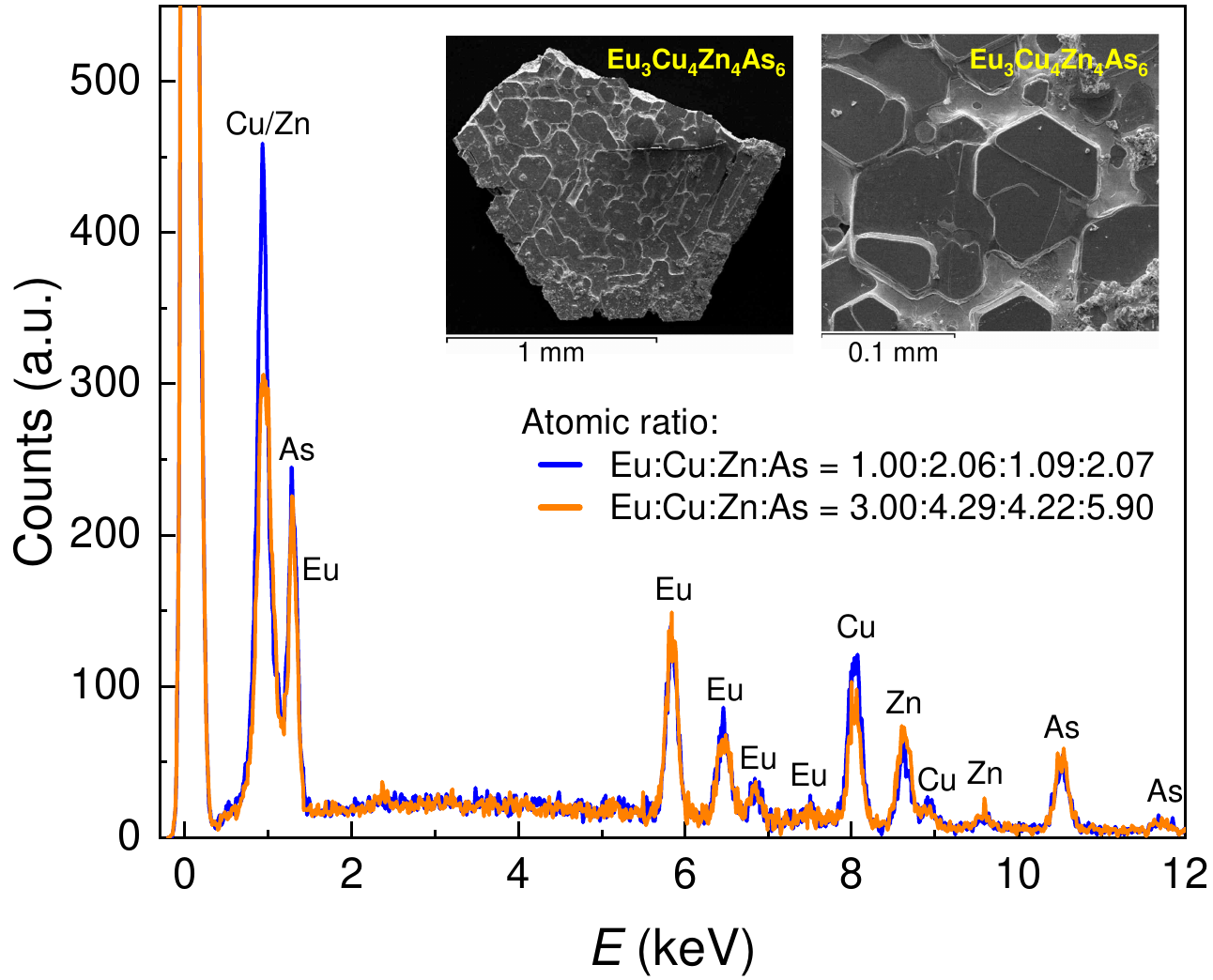}
	\caption{EDS spectra of $\text{EuCu}_2\text{ZnAs}_2$ (blue curve) and $\text{Eu}_3\text{Cu}_4\text{Zn}_4\text{As}_6$ (orange curve), with the normalized atomic percentages listed in the panel. The upper-right insets show representative SEM images of a $\text{Eu}_3\text{Cu}_4\text{Zn}_4\text{As}_6$ crystal, giving a global view and a detailed view, respectively.}
	\label{F2}
\end{figure}

Systematic synthetic exploration showed that only the two $RP = 3$ architectures ($\text{EuCu}_2\text{ZnAs}_2$ and $\text{Eu}_3\text{Cu}_4\text{Zn}_4\text{As}_6$) could be isolated as phase-pure single crystals. SCXRD confirmed that both compounds adopt the trigonal space group $R\bar{3}m$ (Table~\ref{T1}). EDS spectra (Fig.~\ref{F2}) confirm stoichiometric ratios matching the $1:2:1:2$ and $3:4:4:6$ compositions, with $\text{EuCu}_2\text{ZnAs}_2$ showing characteristically stronger Cu peaks because of its higher Cu content. SEM reveals plate-like crystal morphologies with stepped, interconnected hexagonal micro-domains on the surfaces (inset, Fig.~\ref{F2}). This morphology reflects the growth anisotropy and the kinetic difficulty of $c$-axis stacking during alternating block-layer crystallization, which also accounts for minor intergrowth stacking faults and residual Pb flux trapped in surface crevices. These imperfections do not affect the reliability of the structure determination. Although both phases produce a number of Level A and B checkCIF alerts, none reflects a genuine deficiency in the structural model, and each alert is addressed individually in Sec. S7 of the SM \cite{SM}.

Analysis of the structural refinements reveals systematic lattice accommodations upon intergrowth formation (Table~\ref{T1} and Fig.~\ref{F1}). Both intergrowth phases have nearly identical in-plane lattice constants ($a \approx 4.202$~\AA), matching the arithmetic mean of the parent phases. In $\text{EuCu}_2\text{ZnAs}_2$, the $[\text{EuZn}_2\text{As}_2]$ BL contracts along the $c$-axis to $7.100$~\AA\ (versus $7.181$~\AA\ in pure $\text{EuZn}_2\text{As}_2$), whereas the $[\text{EuCu}_4\text{As}_2]$ BL expands to $7.776$~\AA\ (versus $7.718$~\AA\ per BL in pure $\text{EuCu}_4\text{As}_2$). This anisotropic relaxation reflects charge redistribution across the hetero-interface: electrons transfer from the metallic $[\text{EuCu}_4\text{As}_2]$ BLs to the narrow-bandgap Zintl $[\text{EuZn}_2\text{As}_2]$ BLs. Such charge transfer may also come partly from minor $\text{Cu/Zn}$ site mixing across the block layers, despite the distinct valence states and coordination preferences of $\text{Cu}$ and $\text{Zn}$, which could help stabilize the intergrowth structure; this subtle disorder remains unresolvable by standard XRD refinements. The electrostatic changes induced by charge redistribution contract the [$\text{EuZn}_2\text{As}_2$] BLs and expand the [$\text{EuCu}_4\text{As}_2$] BLs along the $c$-axis. A similar trend appears in $\text{Eu}_3\text{Cu}_4\text{Zn}_4\text{As}_6$; however, its $[\text{EuZn}_2\text{As}_2]$ BL thickness ($7.139$~\AA) is less contracted than in $\text{EuCu}_2\text{ZnAs}_2$. The additional Zintl BLs therefore distribute the interfacial strain and transferred charge over a larger volume, reducing the per-layer compression.

The failure to synthesize the $RP = 1$ variant of $\text{EuCu}_2\text{ZnAs}_2$ and the $3:1$ intergrowth $\text{Eu}_2\text{Cu}_6\text{ZnAs}_4$ reveals the thermodynamic and kinetic criteria that govern intergrowth stability. For the $RP = 1$ model, forcing a primitive unit cell disrupts the rhombohedral translation vector $(2/3, 1/3, 1/3)$, creating asymmetric, energetically unfavorable interfacial coordinations that destabilize the structure. The non-formation of high-$n$ intergrowths such as $\text{Eu}_2\text{Cu}_6\text{ZnAs}_4$ ($n=3$, three consecutive $[\text{EuCu}_4\text{As}_2]$ BLs) stems from three mechanisms acting together:
(i) Elastic strain accumulation: thin alternating BLs accommodate lattice mismatch by elastic deformation, but stacking three consecutive identical BLs makes coherency strain build up beyond a critical thickness, triggering dislocation formation or phase separation \cite{matthews1974,zunger1997}.
(ii) Diminished thermodynamic driving force: the thermodynamic stabilization of intergrowth phases comes from interfacial energy reduction driven by charge transfer \cite{hwang2012,Wang2021Blocklayermodelforintergrowthstructures}. As $n$ increases, the interface-to-volume ratio falls, weakening the stabilization energy per volume and causing the system to segregate into parent bulk phases.
(iii) Kinetic stacking faults: nucleating long-period sequences such as $-C_1C_2C_3Z_1-$ carries a high entropic penalty, promoting kinetic ``counting errors'' during growth that favor disordered or multiphase domains over long-range ordered structures \cite{frank1951}. Owing to similar kinetic barriers, we have not observed any higher-order homologues with three consecutive $[\mathrm{EuZn_2As_2}]$ BLs per repeat unit, which would have a sequence such as $-C_1Z_1Z_1Z_1-C_2Z_2Z_2Z_2-C_3Z_3Z_3Z_3-$.

Two structural and thermodynamic principles follow from these outcomes. The first is periodicity inheritance: combining BLs with differing repeat periods makes the intergrowth preferentially inherit the larger repeat period ($RP = 3$) of the parent, preserving global translational symmetry along the stacking axis. The second is interfacial maximization: the system favors a spatially uniform distribution of distinct BLs, such as alternating $1:1$ or $1:2$ sequences, over clustered configurations, because this maximizes the stabilizing interfacial contact area and lowers the free energy.

These principles account for the stability of the intergrowths and their preference for uniform BL distributions, but they do not distinguish the $1:1$ and $1:2$ sequences. That distinction comes from kinetically controlled nucleation within the supersaturated metastable zone of the Pb flux. Under slow cooling ($1\,^{\circ}\mathrm{C/h}$), the system has time to access the nucleation pathway with the lower interfacial energy penalty, which favors the $1:2$ sequence: building $-C_1Z_1Z_1-C_2Z_2Z_2-C_3Z_3Z_3-$ requires the consecutive insertion of two $[\mathrm{EuZn_2As_2}]$ blocks at the same stacking front, a process that is kinetically more demanding than the strict $1:1$ alternation of 1212. Faster cooling ($3\,^{\circ}\mathrm{C/h}$) drives the melt deeper into supersaturation before nucleation, selecting the sequence with the lower nucleation barrier and the simpler step-by-step alternation, i.e., $\mathrm{EuCu_2ZnAs_2}$. This parallels the cooling-rate-controlled phase selection reported for multinary intermetallics grown from metal fluxes, where slow and fast cooling give different products from identical charges without inter-phase conversion.

Taken together, these structural, thermodynamic, and kinetic considerations make the isolated pair, $\text{EuCu}_2\text{ZnAs}_2$ and $\text{Eu}_3\text{Cu}_4\text{Zn}_4\text{As}_6$, a clean system for testing how inserting discrete Zintl spacers tunes interlayer magnetic coupling and transport.

\subsection{Magnetic phase transitions and anisotropic properties}

Both $\text{EuCu}_2\text{ZnAs}_2$ and $\text{Eu}_3\text{Cu}_4\text{Zn}_4\text{As}_6$ show distinct multi-step magnetic transitions that contrast with those of their parent phases ($\text{EuZn}_2\text{As}_2$ and $\text{EuCu}_4\text{As}_2$). The BL stacking sequence sets their transition temperatures and orderings.

\begin{figure*}
	\centering
	\includegraphics[width=0.95\textwidth]{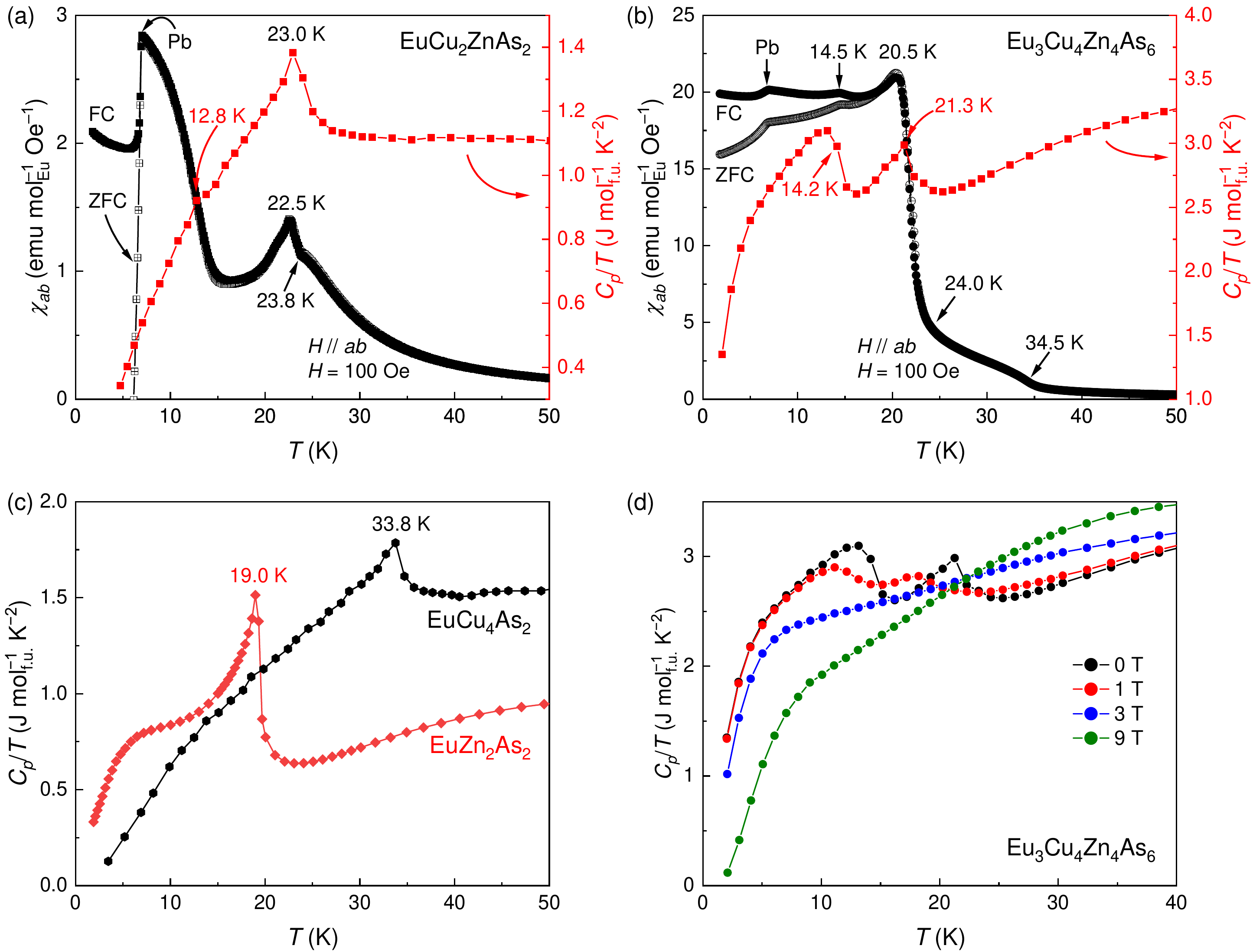}
	\caption{(a, b) Temperature dependence of the in-plane zero-field-cooled (ZFC) and field-cooled (FC) magnetic susceptibility measured under 100 Oe (black curves, left axis), and the zero-field specific heat divided by temperature $C_p/T$ (red curves, right axis) for (a) $\text{EuCu}_2\text{ZnAs}_2$ and (b) $\text{Eu}_3\text{Cu}_4\text{Zn}_4\text{As}_6$. (c) Zero-field $C_p/T$ for $\text{EuCu}_4\text{As}_2$ (black squares) and $\text{EuZn}_2\text{As}_2$ (red hexagons), with data digitized from Refs.~\cite{Li2022Largepositiveandnegativemagnetoresistanceinthemagnetic,Wang2022Anisotropyofthemagneticandtransportproperties}. (d) $C_p/T$ of $\text{Eu}_3\text{Cu}_4\text{Zn}_4\text{As}_6$ under various applied magnetic fields.}
	\label{F3}
\end{figure*}

Figs.~\ref{F3}(a) and \ref{F3}(b) display the in-plane magnetic susceptibility ($\chi_{ab}$, left axis) and the zero-field specific heat divided by temperature ($C_p/T$, right axis) below 50~K for $\text{EuCu}_2\text{ZnAs}_2$ and $\text{Eu}_3\text{Cu}_4\text{Zn}_4\text{As}_6$, respectively. For comparison, the parent phases $\text{EuZn}_2\text{As}_2$ and $\text{EuCu}_4\text{As}_2$ undergo AFM ordering at $T_{\mathrm{N}} = 19$~K \cite{Wang2022Anisotropyofthemagneticandtransportproperties} and FM ordering at $T_{\mathrm{C}} = 34$~K \cite{Li2022Largepositiveandnegativemagnetoresistanceinthemagnetic}, respectively (Fig.~\ref{F3}(c)). The marked differences in transition temperature confirm that the magnetic anomalies in the intergrowths are intrinsic and do not come from parent-phase impurities.

For $\text{EuCu}_2\text{ZnAs}_2$ (Fig.~\ref{F3}(a)), cooling below 50~K produces a continuous rise in $\chi_{ab}$. A slope change appears near 23.8~K, followed by a sharp cusp at $T_{\mathrm{N}} = 22.5$~K that marks long-range AFM order. This transition is confirmed by a $\lambda$-type anomaly in $C_p/T$ at 23.0~K. Below $T_{\mathrm{N}}$, $\chi_{ab}$ decreases gradually before an upturn near 14~K. We attribute this feature to a weak AFM-to-AFM spin-reorientation transition. Two observations support this assignment: (i) the ZFC and FC curves remain overlapped near 14~K without bifurcation, ruling out macroscopic FM hysteresis; and (ii) $C_p/T$ shows only a minute kink around 12.8~K, characteristic of a continuous spin-reorientation process. The abrupt drop in susceptibility below 7~K is caused by the superconducting transition of residual Pb flux on the crystal surface.

$\text{Eu}_3\text{Cu}_4\text{Zn}_4\text{As}_6$ shows richer magnetic behavior (Fig.~\ref{F3}(b)). On cooling from 50~K, $\chi_{ab}$ increases monotonically, with an accelerated rise and a broad hump around 34.5~K attributed to trace $\text{EuCu}_4\text{As}_2$ impurities. Below 24~K, $\chi_{ab}$ rises rapidly, reflecting enhanced FM fluctuations. A pronounced peak at $T_{\mathrm{N}} = 20.5$~K marks the entry into an AFM ordered state. Field-dependent specific-heat measurements confirm the AFM nature of this state (Fig.~\ref{F3}(d)), where the $C_p/T$ peak shifts to lower temperatures under increasing field. The ZFC and FC curves bifurcate below 20.5~K, indicating a small net uncompensated moment, so the ground state of $\text{Eu}_3\text{Cu}_4\text{Zn}_4\text{As}_6$ is a canted AFM structure rather than a simple collinear AFM alignment. This weak FM component originates from the crystal structure: the two crystallographically inequivalent $\text{Eu}^{2+}$ sites (arrows in Fig.~\ref{F1}(a)), sandwiched between alternating $[\text{EuZn}_2\text{As}_2]$ and $[\text{EuCu}_4\text{As}_2]$ BLs, experience distinct local chemical environments, preventing complete antiparallel cancellation of the spins. Because the ZFC--FC splitting sets in only below 20.5~K (well below the $T_{\mathrm{C}} = 34$~K of $\text{EuCu}_4\text{As}_2$), it cannot come from foreign FM impurities. Near 14.5~K, a secondary weak peak in both ZFC and FC curves, with a broad humped anomaly in $C_p/T$, marks a second AFM-like spin-reorientation transition.

\begin{figure*}
	\centering
	\includegraphics[width=0.95\textwidth]{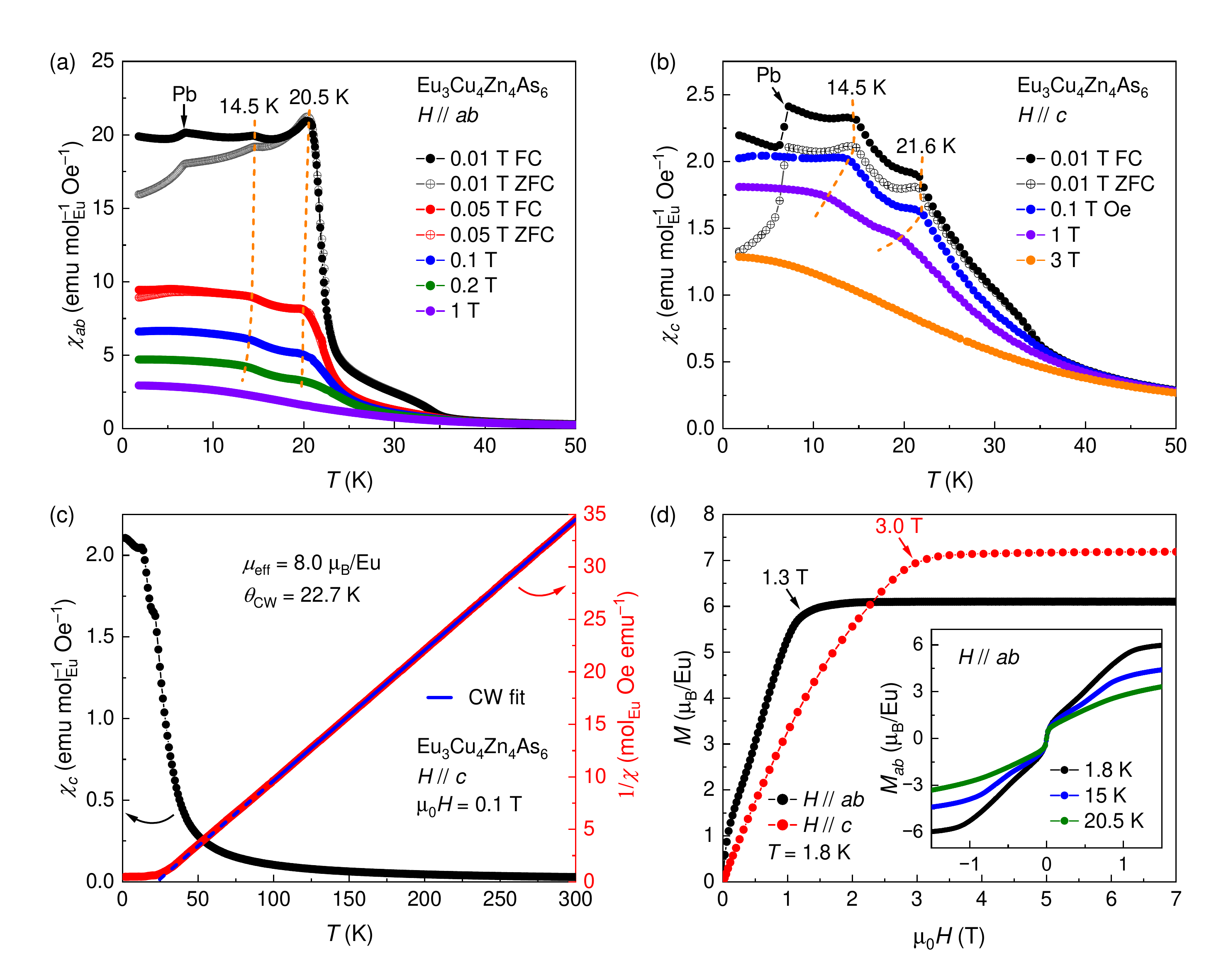}
	\caption{Magnetic properties of $\text{Eu}_3\text{Cu}_4\text{Zn}_4\text{As}_6$ single crystals. (a, b) Temperature dependence of the magnetic susceptibility measured for (a) $H \parallel ab$ and (b) $H \parallel c$. Orange dotted lines mark the characteristic phase-transition peaks to show their field-induced shifts. (c) Curie--Weiss fitting under $\mu_0 H = 0.1$~T, where the blue solid line is the linear fit to the inverse susceptibility ($1/\chi$, right axis). (d) Isothermal magnetization $M(H)$ curves at 1.8~K for $H \parallel ab$ (black) and $H \parallel c$ (red). The inset shows the in-plane $M(H)$ curves at 1.8 K, 15 K, and 20.5 K.}
	\label{F4}
\end{figure*}

The magnetic resemblance between $\text{EuCu}_2\text{ZnAs}_2$ ($T_{\mathrm{N}} = 22.5$~K) and $\text{Eu}_3\text{Cu}_4\text{Zn}_4\text{As}_6$ ($T_{\mathrm{N}} = 20.5$~K) directly reflects their shared constituent BLs. Long-range order in both compounds arises from the competition between the AFM interactions of the $[\text{EuZn}_2\text{As}_2]$ BLs and the FM interactions mediated by the $[\text{EuCu}_4\text{As}_2]$ BLs. In $\text{Eu}_3\text{Cu}_4\text{Zn}_4\text{As}_6$, the $[\text{EuCu}_4\text{As}_2]$ BLs are diluted and separated by thicker $[\text{EuZn}_2\text{As}_2]$ spacers ($1:2$ ratio), so the AFM exchange within the $[\text{EuZn}_2\text{As}_2]$ blocks dominates, and its $T_{\mathrm{N}}$ (20.5~K) stays close to that of pure $\text{EuZn}_2\text{As}_2$ (19~K). Conversely, the $1{:}1$ stacking in $\mathrm{EuCu_2ZnAs_2}$ places a metallic $[\mathrm{EuCu_4As_2}]$ block next to every Zintl $[\mathrm{EuZn_2As_2}]$ block, maximizing the FM--AFM competition at the hetero-interfaces and strengthening the net inter-block exchange. The ordering temperature rises to 22.5 K, a value that, like the 20.5 K of the 3446 phase, lies between the two parents (19 K for AFM $\mathrm{EuZn_2As_2}$ and 34 K for FM $\mathrm{EuCu_4As_2}$). The stacking ratio thus tunes the balance of competing inter-block exchange rather than selecting either parent's order, giving a disorder-free lever for modulating interlayer magnetic coupling.

To characterize the magnetic anisotropy of $\text{Eu}_3\text{Cu}_4\text{Zn}_4\text{As}_6$, field- and orientation-dependent measurements were performed. As shown in Fig.~\ref{F4}(a), an in-plane magnetic field suppresses the $T_{\mathrm{N}} = 20.5$~K peak and shifts it to lower temperatures, as expected for an AFM system. Out-of-plane measurements ($\chi_c$, Fig.~\ref{F4}(b)) reveal pronounced magnetic anisotropy, with $\chi_c$ nearly an order of magnitude smaller than $\chi_{ab}$, indicating that the easy magnetization plane lies within the $ab$ plane. On cooling, $\chi_c$ shows a plateau-like anomaly near 21.6~K with ZFC--FC splitting, and a secondary feature at 14.5~K corresponding to the spin reorientation. A field of only 500~Oe suppresses the ZFC--FC bifurcation in the $ab$ plane, confirming how weak the uncompensated FM moment is.

Fig.~\ref{F4}(c) shows the Curie--Weiss fit of the paramagnetic susceptibility ($100$--$300$~K). The extracted Weiss temperature ($\theta = 22.7$~K) is positive, indicating dominant FM correlations at high temperature. The effective magnetic moment ($\mu_{\mathrm{eff}} = 8.0~\mu_{\mathrm{B}}/\text{Eu}^{2+}$) matches the free-ion value for $\text{Eu}^{2+}$ ($7.94~\mu_{\mathrm{B}}$, $S = 7/2$), confirming the divalent state of Eu.

Isothermal magnetization curves ($M$--$H$) at 1.8~K (Fig.~\ref{F4}(d)) show that the out-of-plane magnetization saturates at $7.0~\mu_{\mathrm{B}}/\text{Eu}^{2+}$ above 3~T, whereas the in-plane magnetization saturates at a lower field of 1.3~T. The slight deviation of the measured in-plane moment ($\sim 6~\mu_{\mathrm{B}}/\text{Eu}^{2+}$) from the theoretical limit comes from the uncertainty in weighing tiny plate-like crystals. The low-field inset reveals a sharp linear magnetization jump below 0.1~T without an open hysteresis loop, consistent with spin-canted weak ferromagnetism. Because this low-field feature persists at 20.5~K, it is distinct from any superconducting flux-trapping effect of residual Pb.

For comparison, the magnetic behavior of $\text{EuCu}_2\text{ZnAs}_2$ is simpler and is given in the SM (Fig. S1) \cite{SM}.

Taken together, the competing inter-block exchange interactions and multi-step transitions in $\text{EuCu}_2\text{ZnAs}_2$ and $\text{Eu}_3\text{Cu}_4\text{Zn}_4\text{As}_6$ show that alternating anion environments tune interlayer magnetic coupling. The spatial modulation of the competing magnetic sublattices breaks local inversion symmetry, which should allow non-trivial spin configurations or unconventional magnetic orders, such as altermagnetism, governed by magnetic point-group symmetries. Determining the microscopic magnetic structures will require neutron diffraction and resonant elastic X-ray scattering (REXS).

\subsection{Magnetotransport properties}

\begin{figure*}
	\centering
	\includegraphics[width=\textwidth]{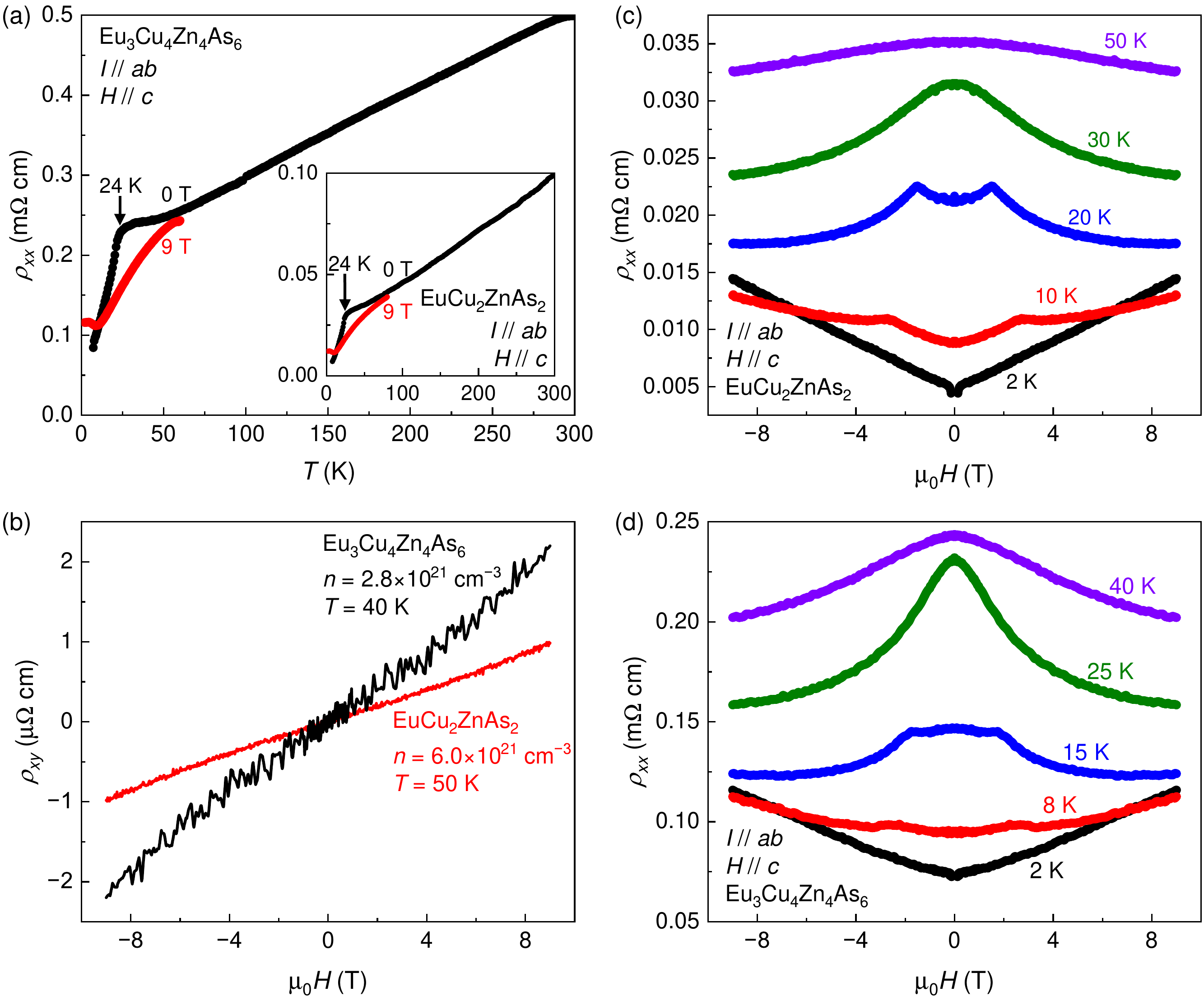}
	\caption{(a) Temperature-dependent electrical resistivity of $\text{Eu}_3\text{Cu}_4\text{Zn}_4\text{As}_6$ at zero field (black) and under 9 T (red). The lower-right inset shows the corresponding data for $\text{EuCu}_2\text{ZnAs}_2$. (b) Hall resistivity of $\text{EuCu}_2\text{ZnAs}_2$ (red) and $\text{Eu}_3\text{Cu}_4\text{Zn}_4\text{As}_6$ (black) in the paramagnetic temperature region. (c, d) Field-dependent resistivity of (c) $\text{EuCu}_2\text{ZnAs}_2$ and (d) $\text{Eu}_3\text{Cu}_4\text{Zn}_4\text{As}_6$ at various temperatures.}
	\label{F5}
\end{figure*}

Fig.~\ref{F5}(a) and its lower-right inset show the temperature-dependent electrical resistivity ($\rho_{xx}$--$T$) of $\text{Eu}_3\text{Cu}_4\text{Zn}_4\text{As}_6$ and $\text{EuCu}_2\text{ZnAs}_2$ at zero field and 9~T. Both compounds are metallic, with a nearly linear decrease in resistivity on cooling that turns into a sharp drop below an inflection near 24~K. This onset matches the enhanced FM fluctuations seen in susceptibility (Figs.~\ref{F3}(a) and \ref{F3}(b)), so short-range FM correlations reduce spin-disorder scattering and raise carrier mobility.

The field-dependent resistivity curves ($\rho_{xx}$--$H$) at various temperatures (Figs.~\ref{F5}(c) and \ref{F5}(d)) show similar MR trends in both materials. At high temperatures ($\ge 25$~K), both compounds show negative MR, reaching $-25\%$ for $\text{EuCu}_2\text{ZnAs}_2$ at 30~K under 9~T and $-30\%$ for $\text{Eu}_3\text{Cu}_4\text{Zn}_4\text{As}_6$ at 25~K under 9~T, driven by field-induced suppression of paramagnetic spin fluctuations. Positive MR dominates at low temperature: at 2~K and 9~T, the resistivity of $\text{EuCu}_2\text{ZnAs}_2$ increases to three times its zero-field value ($\text{MR} \approx +200\%$), whereas that of $\text{Eu}_3\text{Cu}_4\text{Zn}_4\text{As}_6$ increases to 1.6 times ($\text{MR} \approx +60\%$). In the intermediate regime, the $\rho_{xx}(H)$ curves show inflections and plateau-like features. These field-dependent anomalies mark field-induced evolution of the magnetic structure, in which the competing interlayer interactions modulate carrier scattering under field.

The similarity in magnetotransport stems from the shared structural architecture. Because the semiconducting $\text{EuZn}_2\text{As}_2$ has a narrow bandgap and a much lower carrier concentration than the metallic $\text{EuCu}_4\text{As}_2$ \cite{Wang2022Anisotropyofthemagneticandtransportproperties,Li2022Largepositiveandnegativemagnetoresistanceinthemagnetic}, electrical conduction in both intergrowths is dominated by the highly conductive $[\text{EuCu}_4\text{As}_2]$ sheets. Both materials therefore inherit the metallic transport and low-temperature positive MR of the parent $\text{EuCu}_4\text{As}_2$ phase \cite{Li2022Largepositiveandnegativemagnetoresistanceinthemagnetic}.

The main quantitative difference lies in the magnitude of the resistivity: the in-plane resistivity of $\text{Eu}_3\text{Cu}_4\text{Zn}_4\text{As}_6$ is roughly five times higher than that of $\text{EuCu}_2\text{ZnAs}_2$. Hall effect measurements in the paramagnetic regime (Fig.~\ref{F5}(b)) support this. Carrier densities derived from the slope of the Hall resistivity ($\rho_{xy}$--$H$) curves give a carrier concentration for $\text{EuCu}_2\text{ZnAs}_2$ about twice that of $\text{Eu}_3\text{Cu}_4\text{Zn}_4\text{As}_6$. This reflects the higher volumetric stacking density of metallic $[\text{EuCu}_4\text{As}_2]$ BLs in $\text{EuCu}_2\text{ZnAs}_2$ ($1:1$) than in $\text{Eu}_3\text{Cu}_4\text{Zn}_4\text{As}_6$ ($1:2$), where the thicker semiconducting $\text{EuZn}_2\text{As}_2$ spacers dilute the overall carrier density. Full temperature-dependent Hall measurement details are given in the SM (Fig. S4) \cite{SM}.

\section{Conclusion}

By stacking $\text{EuZn}_2\text{As}_2$ and $\text{EuCu}_4\text{As}_2$ parent blocks along the $c$-axis, we designed four intergrowth architectures and synthesized two of them, $\text{EuCu}_2\text{ZnAs}_2$ ($1:1$) and $\text{Eu}_3\text{Cu}_4\text{Zn}_4\text{As}_6$ ($1:2$). Varying the block-layer stacking ratio tunes lattice strain, directional charge transfer, and the physical properties. The repeat period ($RP = 3$) is set by the longer translational symmetry of the Cu-rich parent, and the higher-$n$ members are excluded by accumulating coherency strain and kinetic stacking errors. Inserting additional Zintl BLs modulates interlayer magnetic exchange and dilutes the carrier density, which tunes the transition temperatures and electrical transport.

The microscopic spin arrangements remain to be solved and will require neutron diffraction and REXS. Because $\text{EuZn}_2\text{As}_2$ and $\text{EuCu}_4\text{As}_2$ both belong to large structural families ($\text{CaAl}_2\text{Si}_2$-type and $\text{CaCu}_4\text{P}_2$-type), combining derivatives such as $\text{SrCu}_4\text{As}_2$ with $\text{SrZn}_2\text{As}_2$, or $\text{EuCu}_4\text{As}_2$ with $\text{EuMn}_2\text{As}_2$, should yield many 1212- and 3446-type intergrowths \cite{klufersAB2X2VerbindungenImCaAl2Si2Typ1980,anandMetallicBehaviorInduced2016}. Pairing a parent with exotic electronic states, such as $\text{EuAg}_4\text{Sb}_2$ (incommensurate noncollinear magnetism and giant MR), with a Zintl phase such as $\text{EuCd}_2\text{Sb}_2$ \cite{suMagneticExchangeInduced2020}, would also test how topological electronic states interact with frustrated magnetism. Overall, the block-layer strategy offers a modular, bottom-up route to complex functional materials.

\begin{acknowledgments}
	
This work was supported by the National Natural Science Foundation of China (Grant No. 12674129) and Young Elite Scientists Sponsorship Program by Jiangsu Association for Science and Technology.
\end{acknowledgments}

\section*{DATA AVAILABILITY}
All data supporting the findings of this study are available from the authors upon reasonable request.

\bibliography{reference}
\bibliographystyle{apsrev4-2}
\nocite{REVTEX42Control}

\end{document}



\setcounter{figure}{0}
\setcounter{table}{0}
\setcounter{page}{1}

\title{Supplemental Materials: Stacking-ratio-tuned interlayer coupling in $\text{EuCu}_2\text{ZnAs}_2$ and $\text{Eu}_3\text{Cu}_4\text{Zn}_4\text{As}_6$ intergrowths}

\author{Zhiyu Zhou}
\affiliation{Key Laboratory of Quantum Materials and Devices of Ministry of Education, School of Physics, Southeast University, Nanjing 211189, China}
\author{Xiyu Chen}
\affiliation{Key Laboratory of Quantum Materials and Devices of Ministry of Education, School of Physics, Southeast University, Nanjing 211189, China}
\author{Xin-Yu Zhao}
\affiliation{School of Physics, Interdisciplinary Center for Quantum Information and State Key Laboratory of Silicon and Advanced Semiconductor Materials, Zhejiang University, Hangzhou 310058, China}
\author{Yaran Duan}
\affiliation{Key Laboratory of Quantum Materials and Devices of Ministry of Education, School of Physics, Southeast University, Nanjing 211189, China}
\author{Shiran Tao}
\affiliation{Key Laboratory of Quantum Materials and Devices of Ministry of Education, School of Physics, Southeast University, Nanjing 211189, China}
\author{Guang-Han Cao}
\affiliation{School of Physics, Interdisciplinary Center for Quantum Information, and State Key Laboratory of Silicon and Advanced Semiconductor Materials, Zhejiang University, Hangzhou 310058, China}
\author{Shuai Dong}
\affiliation{Key Laboratory of Quantum Materials and Devices of Ministry of Education, School of Physics, Southeast University, Nanjing 211189, China}
\author{Zhi-Cheng Wang}
\email{wzc@seu.edu.cn}
\affiliation{Key Laboratory of Quantum Materials and Devices of Ministry of Education, School of Physics, Southeast University, Nanjing 211189, China}

\date{\today}

\maketitle

{\footnotesize 
	\begin{center}
		\textbf{Contents}
		\begin{enumerate}
			\item Table S1: Crystallographic data and structural refinement parameters for EuCu$_2$ZnAs$_2$ and Eu$_3$Cu$_4$Zn$_4$As$_6$.
			\item Table S2 \& Table S3: Energy-dispersive X-ray spectroscopy (EDS) compositional data.
			\item Figure S1: Temperature- and field-dependent magnetic properties of EuCu$_2$ZnAs$_2$.
			\item Figure S2: Field-dependent specific heat of EuCu$_2$ZnAs$_2$.
			\item Figure S3: Angle-dependent resistivity anisotropy of Eu$_3$Cu$_4$Zn$_4$As$_6$.
			\item Figure S4: Temperature-dependent Hall resistivity for EuCu$_2$ZnAs$_2$ and Eu$_3$Cu$_4$Zn$_4$As$_6$.
			\item checkCIF/PLATON alerts: Level A and level B alerts and the corresponding point-by-point responses for EuCu$_2$ZnAs$_2$ and Eu$_3$Cu$_4$Zn$_4$As$_6$.
		\end{enumerate}
	\end{center}
}
	\clearpage

	\section{Crystallographic Data and Structure Refinement Parameters}

 \begin{table}[H]
		\footnotesize
		\caption{Crystallographic data and structure refinement parameters for EuCu$_2$ZnAs$_2$ and Eu$_3$Cu$_4$Zn$_4$As$_6$ at 300~K, presented alongside the parent phase parameters for comparison.}
		\label{T1}
		\centering
		\begin{tabular}{lllll}
			\hline
			Chemical formula  & EuCu$_2$ZnAs$_2$ &  Eu$_3$Cu$_4$Zn$_4$As$_6$ & EuCu$_4$As$_2$\textsuperscript{\emph{c}}   & EuZn$_2$As$_2$\textsuperscript{\emph{d}}\\
			\hline
			Crystal system   & Trigonal & Trigonal & Trigonal & Trigonal \\
			Space group & $R\bar{3}m$ (No. 166) & $R\bar{3}m$ (No. 166) & $R\bar{3}m$ (No. 166) & $P\bar{3}m1$ (No. 164)\\
			$a$ (\AA)    & 4.2024(4)  & 4.2028(1) & 4.183(1)  & 4.21118(3) \\
			$c$ (\AA)   & 44.626(5) & 66.144(3) & 23.154(2) & 7.18114(6) \\
			$V$ (\AA$^3$) & 682.52(15) & 1011.81(6)  & 350.86(5)  & 110.2888(24) \\
			$Z$ & 6 & 3 & 3 & 1 \\
			Formula weight (g/mol) & 494.10 & 1421.04 & 555.99 & 432.57 \\
			$\rho_\mathrm{calc}$ (g/cm$^3$)  &  7.215 & 6.996 & 7.894 & 6.51 \\
			Temperature (K)  & 300(2)  & 298(2) &  &  \\
			Radiation  & Mo $K\alpha$ & Mo $K\alpha$ &  &  \\
			2$\theta$ range ($^\circ$) & 5.478 $\rightarrow$ 64.202  & 3.694 $\rightarrow$ 57.926 &  &  \\
			$h$ & $-6 \rightarrow  6$ & $-5 \rightarrow  56$ &  &  \\
			$k$ & $-6   \rightarrow   5$  & $-4 \rightarrow  5$ &  &  \\
			$l$ & $-52   \rightarrow  59$  & $-87 \rightarrow  75$ &  &  \\
			Reflections collected  & 3005  & 3113 &  &  \\
			Independent reflections  & 336  & 406 &  &  \\
			$R\mathrm{_{int}}$ & 0.0381  & 0.0429 &  &  \\
			Goodness-of-fit  & 1.240  & 1.097 &  &  \\
			$R_1$ [$F^2>2\sigma(F^2)$]\textsuperscript{\emph{a}} & 3.21\%  & 4.83\% &&\\
			$wR_2$ [in units of $F^2$]\textsuperscript{\emph{b}} & 8.58\%  & 11.53\% &&\\
			\hline
		\end{tabular}
		\begin{flushleft}
			\footnotesize{\textsuperscript{\emph{a}} $R_1 = \sum ||F_\mathrm{o}| - |F_\mathrm{c}|| / \sum |F_\mathrm{o}|$.	\textsuperscript{\emph{b}} $wR_2 = [\sum w(F_\mathrm{o}^2 - F_\mathrm{c}^2)^2 / \sum w(F_\mathrm{o}^2)^2]^{1/2}$.

			\textsuperscript{\emph{c}} D{\"u}nner, J.; Mewis, A. Synthesis and crystal structure of ACu$_4$As$_2$ (A: Ca--Ba, Eu). \textit{Z. Anorg. Allg. Chem.} \textbf{1999}, \textit{625}, 625--630.

			\textsuperscript{\emph{d}} Wang, Z.-C.; Been, E.; Gaudet, J.; Alqasseri, G. M. A.; Fruhling, K.; Yao, X.; Stuhr, U.; Zhu, Q.; Ren, Z.; Cui, Y.; Jia, C.; Moritz, B.; Chowdhury, S.; Devereaux, T.; Tafti, F. Anisotropy of the magnetic and transport properties of EuZn$_2$As$_2$. \textit{Phys. Rev. B} \textbf{2022}, \textit{105}, 165122.
			}
		\end{flushleft}

	\end{table}

	\clearpage

	\section{Energy-Dispersive X-Ray Spectroscopy (EDS) Data}

	\begin{table}[htbp]
		\caption{Chemical compositions determined by EDS for seven single-crystal regions of EuCu$_2$ZnAs$_2$, normalized to Eu = 1.00.}
		\label{T2}
		\centering
		\begin{tabular}{lllllllll}
			\hline
			Element  & S1 & S2 & S3 & S4 & S5 & S6 & S7 & Avg. \\
			\hline

			Eu  & 1.00 &	1.00 &	1.00 &	1.00 &	1.00 &	1.00 &	1.00 &	1.00 \\
			Cu  & 1.95 &	2.03 &	1.99 &	2.05 &	2.20 &	2.40 &	1.99 &	2.09 \\
			Zn  & 0.99 &	1.05 &	0.93 &	1.09 &	1.06 &	0.94 &	0.93 &	1.00 \\
			As  & 1.87 &	2.29 &	2.17 &	2.07 &	2.49 &	2.00 &	2.17 &	2.15 \\

			\hline
		\end{tabular}
	\end{table}

	\begin{table}[htbp]
		\caption{Chemical compositions determined by EDS for seven single-crystal regions of Eu$_3$Cu$_4$Zn$_4$As$_6$, normalized to Eu = 3.00.}
		\label{T3}
		\centering
		\begin{tabular}{lllllllll}
			\hline
			Element  & S1 & S2 & S3 & S4 & S5 & S6 & S7 & Avg. \\
			\hline
			Eu  & 3.00 &	3.00 &	3.00 &	3.00 &	3.00 &	3.00 &	3.00 &	3.00 \\
			Cu  & 4.26 &	3.74 &	3.88 &	4.15 &	4.08 &	4.01 &	3.63 &	3.96 \\
			Zn  & 4.27 &	3.98 &	4.07 &	4.07 &	4.09 &	4.21 &	3.60 &	4.04 \\
			As  & 7.11 &	7.06 &	6.82 &	7.03 &	6.90 &	6.90 &	5.68 &	6.78 \\

			\hline
		\end{tabular}
	\end{table}

	\clearpage

	\section{Magnetic Characterization of EuCu$_2$ZnAs$_2$}

	\begin{figure}[htbp]
		\centering
		\includegraphics[width=\columnwidth]{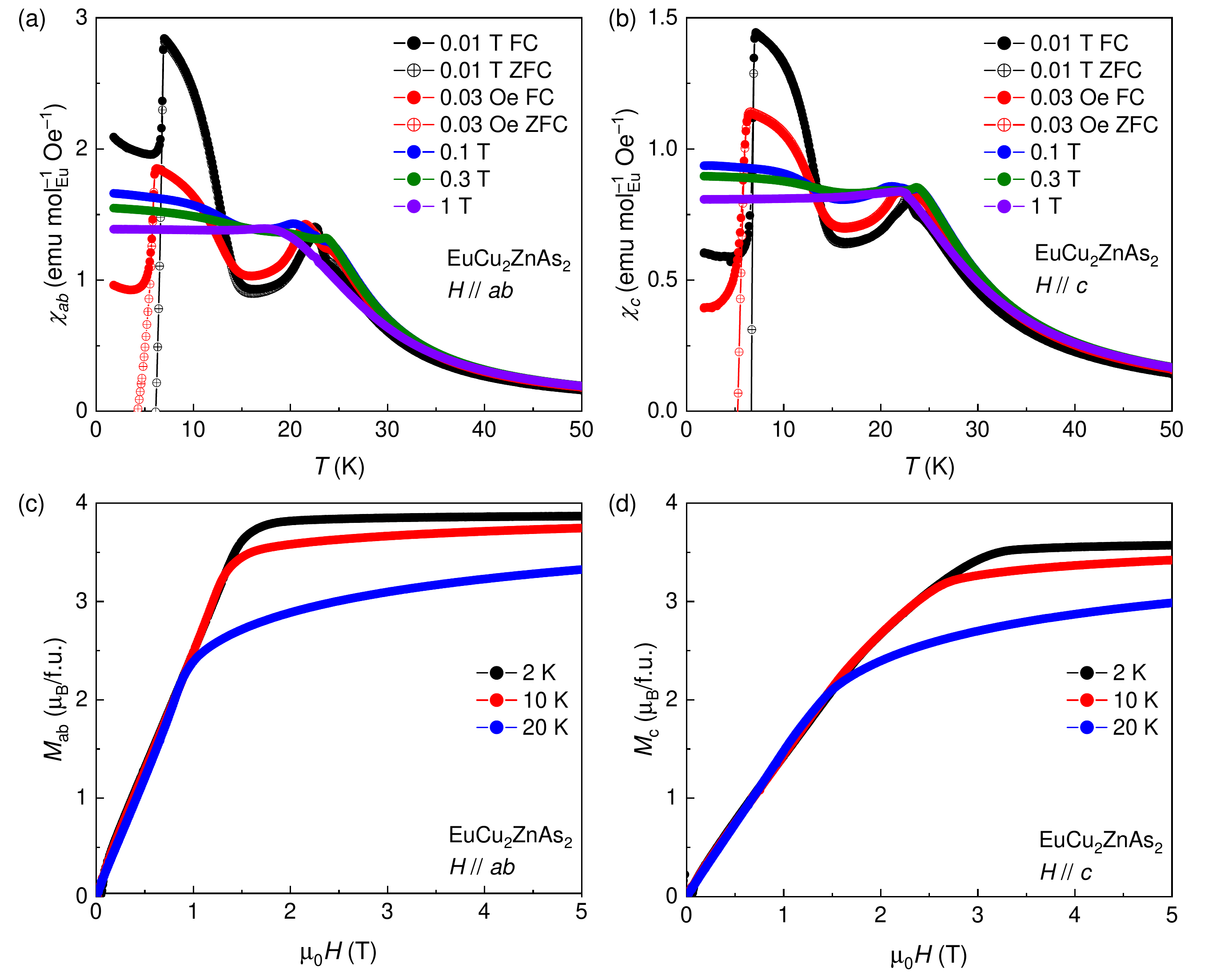}
		\caption{Magnetic characterization of single-crystalline EuCu$_2$ZnAs$_2$. (a, b) Temperature dependence of magnetic susceptibility ($\chi$) measured under $H \parallel ab$ (a) and $H \parallel c$ (b). (c, d) Isothermal magnetization curves ($M$--$H$) recorded at 1.8~K for $H \parallel ab$ (c) and $H \parallel c$ (d). The observed saturation magnetization is slightly lower than the theoretical value of $7~\mu_\mathrm{B}/\text{Eu}^{2+}$ because of the residual Pb flux on the crystal surfaces.}
		\label{fig:Fig-S1}
	\end{figure}

	\clearpage

	\section{Specific Heat of EuCu$_2$ZnAs$_2$}

	\begin{figure}[htbp]
		\centering
		\includegraphics[width=0.85\columnwidth]{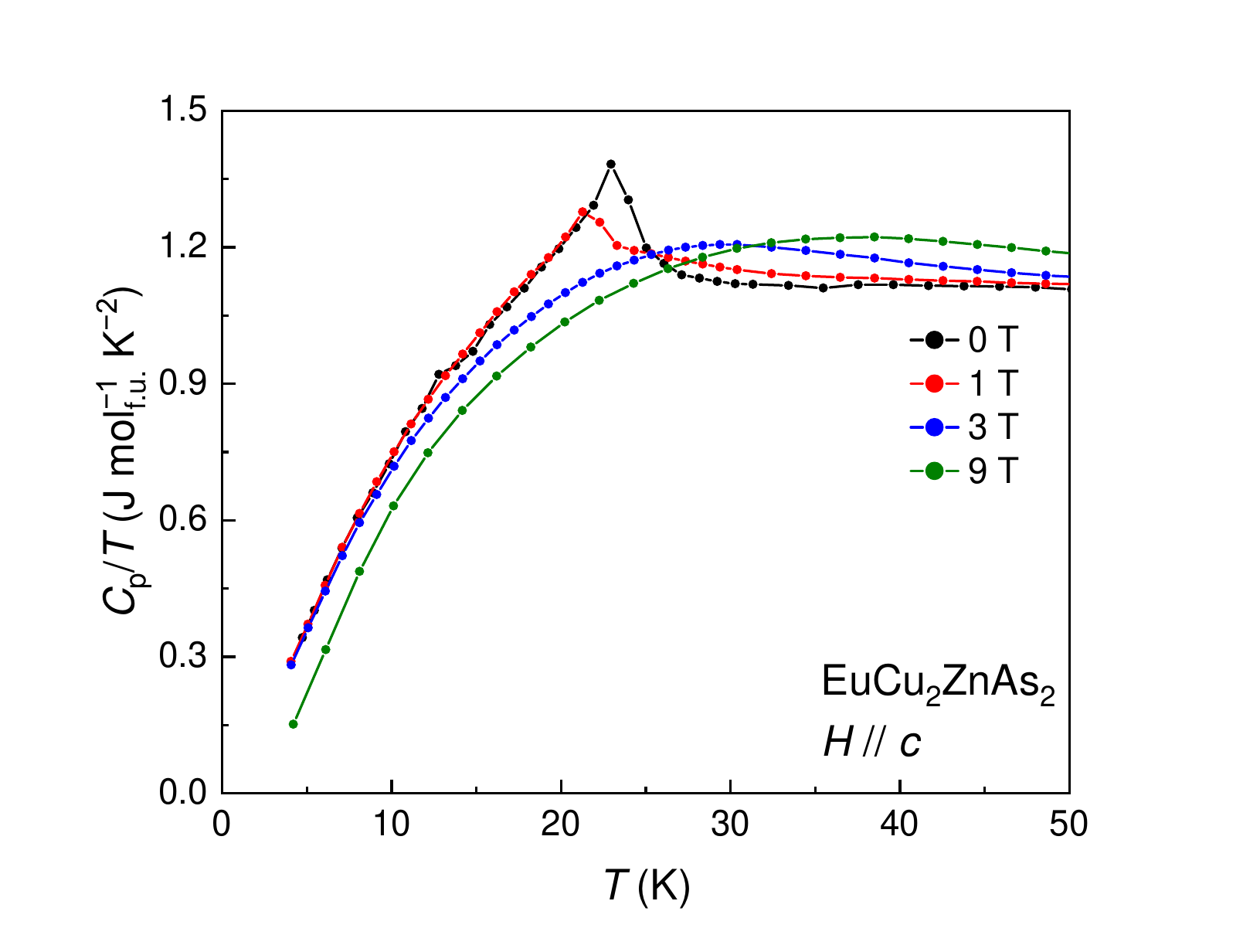}
		\caption{Temperature-dependent specific heat divided by temperature ($C_p/T$) of single-crystalline EuCu$_2$ZnAs$_2$ under various applied magnetic fields.}
		\label{fig:Fig-S2}
	\end{figure}

	\clearpage

	\section{Resistivity Anisotropy of Eu$_3$Cu$_4$Zn$_4$As$_6$}

	\begin{figure}[htbp]
		\centering
		\includegraphics[width=0.85\columnwidth]{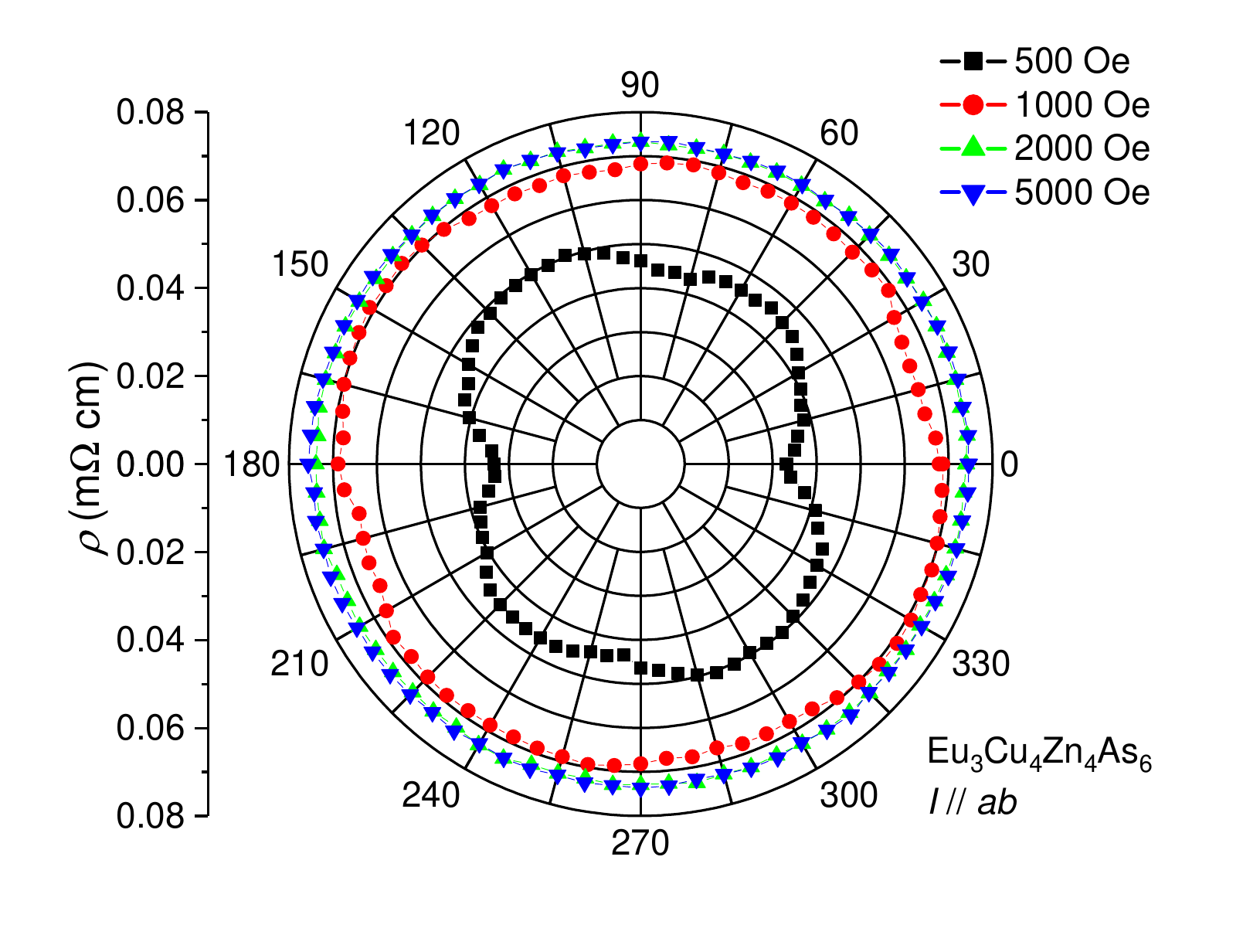}
		\caption{Angular dependence of resistivity for Eu$_3$Cu$_4$Zn$_4$As$_6$ under several magnetic fields, showing a weak magnetoresistance anisotropy with respect to the field orientation. Here, $\theta = 0^\circ$ and $90^\circ$ correspond to $H \perp ab$ ($H \parallel c$) and $H \parallel ab$, respectively.}
		\label{fig:Fig-S3}
	\end{figure}

	\clearpage

	\section{Hall Effect Measurements}

	\begin{figure}[htbp]
		\centering
		\includegraphics[width=\columnwidth]{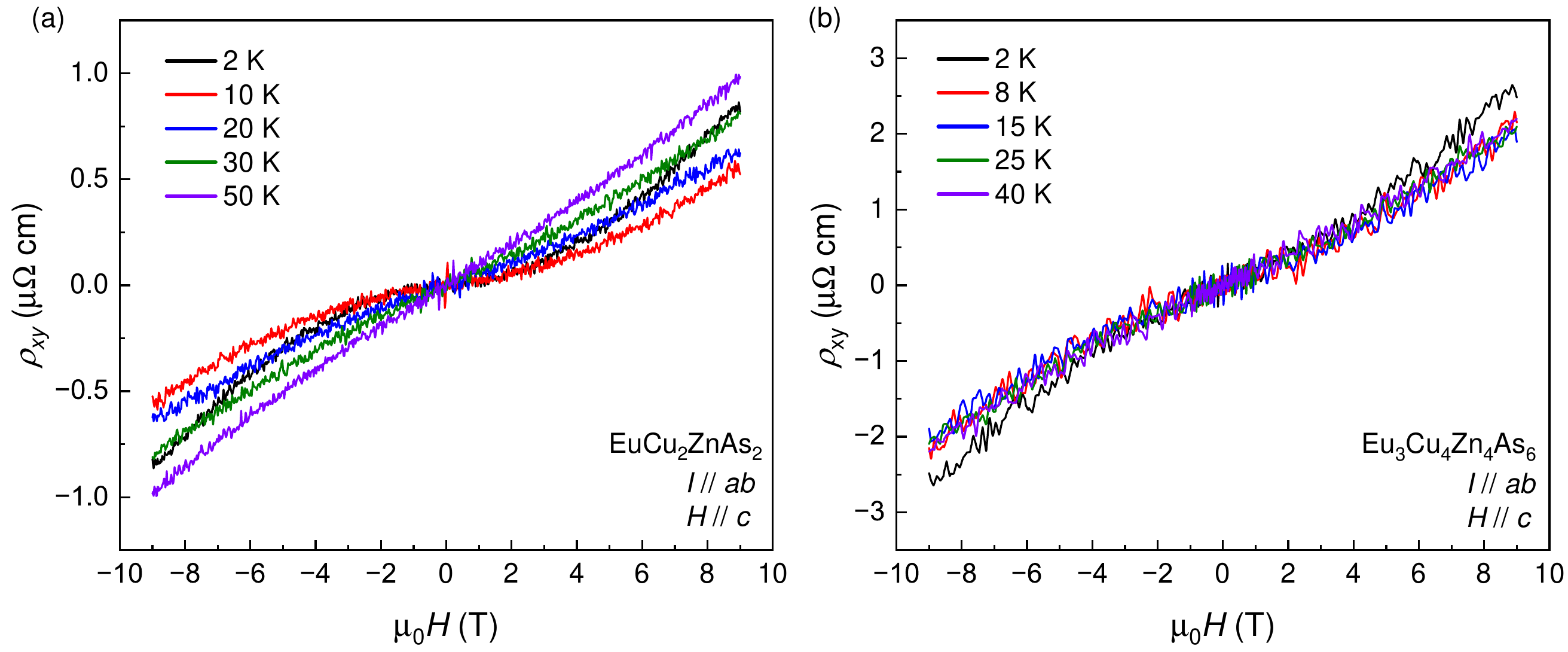}
		\caption{Hall resistivity ($\rho_{xy}$) as a function of magnetic field for EuCu$_2$ZnAs$_2$ and Eu$_3$Cu$_4$Zn$_4$As$_6$ measured at representative temperatures.}
		\label{fig:Fig-S4}
	\end{figure}

	\clearpage

	\section{checkCIF/PLATON Alerts and Point-by-Point Responses}

	Both structures were validated with the IUCr checkCIF/PLATON service and the
	resulting reports were deposited together with the corresponding CIFs. Below
	we reproduce the complete set of \textbf{level A and level B} alerts (the
	only alert levels that require a response) and address them one by one. No
	level A or level B alert concerns the structural model itself: every one of
	them refers to residual electron density in the final difference Fourier map,
	and every extremum lies on, or immediately adjacent to, an atom position
	that is already present in the model. Neither compound gives an alert on the
	space group, the atom connectivity, the displacement parameters, or the
	internal consistency of the reflection data.

	\subsection{EuCu$_2$ZnAs$_2$ (1212 phase)}

	The refinement converged with SHELXL-2019/3 against all 336 unique
	reflections ($R_1 = 3.21\%$, $wR_2 = 8.58\%$, GooF $= 1.240$, 23
	parameters), with all atoms refined anisotropically and all displacement
	parameters positive definite. The difference map ranges from $+3.36$ to
	$-3.78$~e\,\AA$^{-3}$, and every extremum falls within $\sim$1.8~\AA{} of a
	metal position of the model. The alerts are:

	{\small
	\begin{verbatim}
	Alert level A
	PLAT972_ALERT_2_A Check Calcd Resid. Dens.  1.74Ang From Zn04    .  -3.54 eA-3

	Alert level B
	PLAT971_ALERT_2_B Check Calcd Resid. Dens.  0.84Ang From Eu01    .   2.56 eA-3
	PLAT971_ALERT_2_B Check Calcd Resid. Dens.  0.86Ang From Eu01    .   2.51 eA-3
	PLAT972_ALERT_2_B Check Calcd Resid. Dens.  1.52Ang From Cu06    .  -3.09 eA-3
	\end{verbatim}
	}

	\subsubsection*{Level A: PLAT972 --- negative residual density near Zn04}

	\emph{Alert.} A negative residual density of $-3.54$~e\,\AA$^{-3}$ is
	located 1.74~\AA{} from Zn04.

	\emph{Response.} This is the largest single feature of the difference map and
	corresponds to a shallow negative ripple adjacent to the Zn04 site, i.e.\ in
	the immediate vicinity of an atom already present in the model. No additional
	atom position is indicated. Together with the smaller extrema listed below,
	such a feature is characteristic of truncation and absorption ripples around
	the strongly scattering metal atoms, and it arises here from the same
	data-level effects as in the related compound Eu$_3$Cu$_4$Zn$_4$As$_6$: (i)
	the layered structure has a long $c$ axis (44.63~\AA) and exhibits stacking
	faults that attenuate the $00l$ reflections, and (ii) the absorption
	coefficient is very high ($\mu = 42.1$~mm$^{-1}$, Mo\,$K\alpha$). The
	averaged structure itself is unaffected and remains well defined ($R_1 =
	3.21\%$, $wR_2 = 8.58\%$); no revision of the model is warranted.

	\subsubsection*{Level B: PLAT971 --- positive residual density near Eu01}

	\emph{Alert.} Two positive residual-density maxima of $+2.56$ and
	$+2.51$~e\,\AA$^{-3}$ are located 0.84 and 0.86~\AA{} from Eu01.

	\emph{Response.} These two features lie immediately adjacent to the Eu01 site
	and are the positive counterpart of the absorption/truncation ripple
	surrounding this strongly scattering europium atom. They do not correspond to
	any missing atom and are of the same origin as the residual density discussed
	above.

	\subsubsection*{Level B: PLAT972 --- negative residual density near Cu06}

	\emph{Alert.} A negative residual density of $-3.09$~e\,\AA$^{-3}$ is
	located 1.52~\AA{} from Cu06.

	\emph{Response.} This negative feature is the companion undershoot of the
	ripple around Cu06 and, like the Zn04 feature above, carries no
	crystallographic significance. It is fully consistent with the truncation and
	absorption effects already described.

	\subsection{Eu$_3$Cu$_4$Zn$_4$As$_6$ (3446 phase)}

	The refinement converged with SHELXL-2019/3 against all 406 unique
	reflections ($R_1 = 4.83\%$, $wR_2 = 11.53\%$, GooF $= 1.097$, 30
	parameters), with all atoms refined anisotropically and all displacement
	parameters positive definite. The largest positive peak of the difference map
	is $+8.62$~e\,\AA$^{-3}$ and the largest negative one about
	$-2.8$~e\,\AA$^{-3}$ (PLAT094 reports a maximum-to-minimum ratio of 3.05).
	The alerts are:

	{\small
	\begin{verbatim}
	Alert level A
	PLAT971_ALERT_2_A Check Calcd Resid. Dens.  1.55Ang From As05    .   9.75 eA-3
	PLAT971_ALERT_2_A Check Calcd Resid. Dens.  0.20Ang From As04    .   6.66 eA-3
	PLAT971_ALERT_2_A Check Calcd Resid. Dens.  0.14Ang From As03    .   5.23 eA-3
	PLAT971_ALERT_2_A Check Calcd Resid. Dens.  1.32Ang From As03    .   4.48 eA-3
	PLAT971_ALERT_2_A Check Calcd Resid. Dens.  1.45Ang From Eu02    .   4.38 eA-3
	PLAT971_ALERT_2_A Check Calcd Resid. Dens.  0.01Ang From As05    .   4.09 eA-3
	PLAT973_ALERT_2_A Check Calcd Positive Resid. Density on      Cu08   8.46 eA-3
	PLAT973_ALERT_2_A Check Calcd Positive Resid. Density on      Eu01   8.38 eA-3
	PLAT973_ALERT_2_A Check Calcd Positive Resid. Density on      Eu02   7.43 eA-3
	PLAT973_ALERT_2_A Check Calcd Positive Resid. Density on      Zn06   5.36 eA-3
	PLAT973_ALERT_2_A Check Calcd Positive Resid. Density on      Zn07   4.35 eA-3
	PLAT973_ALERT_2_A Check Calcd Positive Resid. Density on      Cu09   2.51 eA-3

	Alert level B
	PLAT097_ALERT_2_B Large Reported Max. (Positive) Residual Density  8.62 eA-3
	PLAT972_ALERT_2_B Check Calcd Resid. Dens.  0.68Ang From Eu02    .  -2.55 eA-3
	\end{verbatim}
	}

	\subsubsection*{Level A: PLAT971 and PLAT973 --- maximum residual density}

	\emph{Alerts.} PLAT971\_2\_A lists six positive maxima located 0.01--1.55~\AA{}
	from As or Eu ($+9.75$, $+6.66$, $+5.23$, $+4.48$, $+4.38$ and
	$+4.09$~e\,\AA$^{-3}$); PLAT973\_2\_A lists six maxima located on Cu, Eu or
	Zn ($+8.46$, $+8.38$, $+7.43$, $+5.36$, $+4.35$ and
	$+2.51$~e\,\AA$^{-3}$).

	\emph{Response.} These twelve alerts describe one and the same phenomenon, and
	none of them indicates an additional atom site: every peak lies within
	$\sim$1.6~\AA{} of a metal atom and coincides with a site already present in
	the model. Such a distribution (positive maxima on the heavy-atom cores with
	weaker negative undershoots at the same positions) is the signature of
	truncation and absorption ripples in the difference map of a strongly
	scattering structure, and two features of this sample make it expected here.

	\begin{itemize}
	\item \emph{Stacking disorder.} The compound is a layered polytype built from
	slabs stacked along a very long $c$ axis (66.14~\AA). Stacking faults,
	intrinsic to this architecture, systematically attenuate the $00l$
	reflections, so that a single averaged three-dimensional model in
	$R\bar{3}m$ cannot reproduce the measured $00l$ intensities. The resulting
	mismatch appears as residual density along the stacking direction and within
	the heavy-atom layers, and cannot be removed by adding parameters to an
	ordered model.
	\item \emph{High absorption.} With $\mu = 41.29$~mm$^{-1}$ (Mo\,$K\alpha$)
	and $T_{\mathrm{min}} = 0.310$, residual absorption errors surviving the
	multi-scan correction produce systematic ripples close to the strongest
	scatterers (Eu, As), exactly as observed.
	\end{itemize}

	The averaged structure is nevertheless well determined: all displacement
	parameters are positive definite, the largest shift/esd is 0.009, and the
	agreement factors are $R_1 = 4.83\%$ and $wR_2 = 11.53\%$. We therefore
	retain the ordered average model and attribute the residual density to the
	data-level effects described above.

	\subsubsection*{Level B: PLAT097 and PLAT972 --- maximum positive and negative residual density}

	\emph{Alerts.} A maximum positive residual density of $8.62$~e\,\AA$^{-3}$ is
	reported (PLAT097), and a negative residual density of $-2.55$~e\,\AA$^{-3}$
	is located 0.68~\AA{} from Eu02 (PLAT972).

	\emph{Response.} PLAT097 is the report-level statement of the same density
	discussed above and is accounted for by stacking disorder along $c$ combined
	with the high absorption coefficient. The single negative feature near Eu02 is
	the undershoot that accompanies the positive ripple surrounding this site; the
	positive and negative extrema of the map are both localized at heavy-atom
	positions and share the same origin. Neither carries crystallographic
	significance, and no revision of the model is warranted.

	\subsection*{Summary}

	All level A and level B alerts of both compounds arise from residual electron
	density localized at the heavy-atom sites and none of them requires a change
	of the structural model. The larger amplitudes in
	Eu$_3$Cu$_4$Zn$_4$As$_6$ reflect its longer $c$ axis
	(66.14~\AA), and hence the stronger $00l$ attenuation, as well as the
	absorption ripples around Eu and As; EuCu$_2$ZnAs$_2$ shows the
	same pattern with a threefold smaller amplitude ($+3.36$ to
	$-3.78$~e\,\AA$^{-3}$). The ordered anisotropic models thus provide a
	well-defined description of both structures.